\documentclass{kapedbk}
\usepackage{graphicx,amsmath,amsfonts,edbkps,amsbsy,amssymb,bm,color}
\normallatexbib

\renewcommand{\Re}{\mathop{\mathrm{Re}}}

\newcommand{\av}[1]{\langle #1 \rangle}
\newcommand{\trace}{\mathop{\mathrm{Tr}}}
\newcommand{\f}{fluctuator }
\newcommand{\fs}{fluctuators }

\newcommand{\ba}{\mathbf {a}}

\newcommand{\bn}{\mathbf {n}}
\newcommand{\bB}{\mathbf {B}}
\newcommand{\bF}{\mathbf {F}}

\newcommand{\br}{\mathbf {r}}

\newcommand{\cP}{\mathcal {P}}
\newcommand{\cH}{\mathcal {H}}
\newcommand{\cK}{\mathcal {K}}

\newcommand{\cN}{\mathcal {N}}
\newcommand{\cS}{\mathcal {S}}

\newcommand{\cX}{\mathcal {X}}

\newcommand{\prl}{Phys. Rev. Lett.\, }
\newcommand{\prb}{Phys. Rev. B\, }
\newcommand{\rmp}{Rev. Mod. Phys.\, }
\newcommand{\onlinecite}{\cite}

\begin{document}
\articletitle{Low-frequency noise as a source\\ of dephasing of a qubit}

\chaptitlerunninghead{Dephasing of a qubit by $1/f$-noise}
\author{Y. M. Galperin}
\affil{Department of Physics, University of Oslo, PO Box 1048
  Blindern, 0316 Oslo, Norway,\\
A. F. Ioffe  Physico-Technical Institute, 194021
  St. Petersburg, Russia, and Argonne National Laboratory, 9700
  S. Cass av., Argonne, 
  IL 60439, USA}
\email{iouri.galperine@fys.uio.no}
\author{B. L. Altshuler}
\affil{Physics Department, Princeton University,
        Princeton, NJ 08544, USA,\\
and NEC Research Institute, 4 Independence Way,
        Princeton, NJ 08540, USA}
\email{bla@feynman.princeton.edu}
\and 
\author{D. V. Shantsev}
\affil{Department of Physics, University of Oslo, PO Box 1048
  Blindern, 0316 Oslo, Norway,\\
and A. F. Ioffe  Physico-Technical Institute, 194021
  St. Petersburg, Russia}
\email{daniil.shantsev@fys.uio.no}

\begin{abstract}
With the growing efforts in isolating solid-state qubits from
external decoherence sources, the material-inherent sources of
noise start to play crucial role. One representative example is
electron traps in the device material or substrate. Electrons can
tunnel or hop between a charged and an empty trap, or between a
trap and a gate electrode. A single trap typically produces
telegraph noise and can hence be modeled as a bistable fluctuator.
Since the distribution of hopping rates is exponentially broad,
many traps produce flicker-noise with spectrum close to $1/f$.
Here we develop a theory of decoherence of a qubit in the
environment consisting of two-state fluctuators, which experience
transitions between their states induced by interaction with
thermal bath. Due to interaction with the qubit the fluctuators
produce $1/f$-noise in the qubit's eigenfrequency. We calculate
the results of qubit manipulations - free induction and echo
signals - in such environment. The main problem is that in many
important cases the relevant random process is both non-Markovian
and non-Gaussian. Consequently the results in general cannot be
represented by pair correlation function of the qubit
eigenfrequency fluctuations. Our calculations are based on
analysis of the density matrix of the qubit using methods
developed for stochastic differential equations. The proper
generating functional is then averaged over different fluctuators
using the so-called Holtsmark procedure. The analytical results
are compared with simulations allowing checking accuracy of the
averaging procedure and evaluating mesoscopic fluctuations. The
results allow understanding some observed features of the echo
decay in Josephson qubits.
\end{abstract}

\section{Introduction and model}
The dynamics of quantum two-level systems has recently attracted
special attention in connection with ideas of quantum computation.
A crucial requirement is to the phase coherence in the presence of
noisy environment~\cite{NiChu}. Solid state devices have many
advantages for realization of quantum computation that has been
confirmed by several successful experiments, for a review see, e.
g., Ref.~\onlinecite{Nak} and references therein.
In solid-state realizations of quantum bits (qubits) the major
intrinsic noise is due to material-specific fluctuations
(substrate, etc). Concrete mechanisms of these fluctuations depend
upon the realization. In particular, in the case of charge qubits
the background  charge fluctuations with $1/f$ spectrum are
considered as most important~\cite{Nak}. They are usually
attributed to random motion of
 charges either between localized impurity states, or between localized
impurity states and metallic electrodes.

The conventional way to allow for  the noisy environment is to
describe it as a set of harmonics oscillators with a certain
frequency spectrum. The resulting ``spin-boson models'' were
extensively discussed in the literature, see for a review
Refs.~\onlinecite{Leggett} and~\onlinecite{Weiss}. Applications of
these models to concrete qubit implementations have been recently
reviewed by Shnirman \emph{et al.} in Ref.~\onlinecite{Shnirman}.

\subsection{Spin-boson model}
Conventionally, the quantum system which we will call the qubit is assumed
to be coupled linearly to an oscillator bath with interaction
Hamiltonian
\begin{equation}
  \label{eq:200}
  \cH =\sigma_z \hat{\cX}\, , \quad  \hat{\cX}= \sum_j C_j
\left(\hat{b}_j+\hat{b}_j^\dag\right)\, .
\end{equation}
Here $\sigma_i$, $i=x,y,z$,  are the Pauli matrices describing the qubit,  while $\hat{b}_j$ and
$\hat{b}_j^\dag$ stand for bosons. The decoherence is then expressed
in terms of the symmetric correlation function
\begin{equation}
  \label{eq:201}
  S_{\cX}(\omega )\equiv \left \langle\left[\hat{\cX}(t),\hat{\cX}(0)
  \right]_+ \right\rangle_{\! \omega } =2J(\omega)\coth \frac{\omega
  }{2T} \, ,
\end{equation}
where $J(\omega )$ the bath spectral density,
 $J(\omega )\equiv \pi \sum_j C_j^2\, \delta (\omega -\omega_j)$.
 Here and below we put $\hbar =1$ and $k_{\text{B}}=1$.
In the simplest case the decoherence can be characterized
by~\cite{Shnirman}
\begin{equation}
\label{eq:203}
\cK(t)=-\ln \trace{\left(e^{-i
\hat{\Phi}(t)}\, e^{i\hat{\Phi}(0)} \hat{\rho}_{b}\right)}
\end{equation}
where $\hat{\rho}_b$ is the density matrix of the thermal bath,
and the bath phase operator is defined as
\begin{equation}
\label{bathphase}
\hat{\Phi}(t) \equiv i \sum_j (2C_j/\omega_j)\,
e^{i\cH_b t} \left(\hat{b}_j^\dag
  -\hat{b}_j\right)e^{-i\cH_b t},
\end{equation}
$\cH_b$ being the bath Hamiltonian. The quantity $\cK(t)$ is
conventionally expressed through 
the bath spectral density, $J(\omega)$, as~\cite{Leggett}
\begin{equation}
  \label{eq:204}
  \cK(t)=\frac{8}{\pi} \int_0^\infty \! \!  \frac{ d \omega\,  J(\omega
  )}{\omega ^2} \left[\sin^2\frac{\omega t}{2}\coth \frac{\omega }{2T} +
  \frac{i}{2}\sin \omega t \right].
 \end{equation}
The most popular assumption about $J(\omega)$, namely
$J(\omega )=(\alpha \pi \omega /2) \Theta (\omega_c -\omega)$, is
``Ohmic dissipation''.
Here $\alpha$ is a dimensionless coupling strength, while $\Theta(x)$
is the Heaviside unit step function.

Important features of this approach is that (i) the decoherence is
determined solely by the \emph{pair correlation function}
$S_{\cX}(\omega )$ that assumes the noise to be Gaussian; (ii)
$S_{\cX}(\omega )$ is related to the bath spectral density through
the \emph{fluctuation-dissipation theorem}, which assumes the
system to be equilibrium.  As long as these assumptions hold,
the method provides powerful tools to analyze the decoherence.

Several attempts, see Ref.~\onlinecite{Shnirman} and
references therein, were made to extend the spin-boson
model to the so-called ``sub-Ohmic'' case, in particular, to the case
of $1/f$-noise where $S_{\cX}(\omega) \propto |\omega|^{-1}$.

We believe that $1/f$ noise is a typical nonequilibrium phenomenon. It is
due to the fact that some excitations of the environment relax so
slowly that  cannot reach the equilibrium  during the
measuring time. As a result, the fluctuation-dissipation theorem
cannot be applied. Moreover, $1/f$ noise is not a stationary Markov process.
Indeed,  it is created by the fluctuators with
exponential broad distribution of the relaxation rates
and thus is not fully characterized by its
pair correlation function.

\subsection{Spin-fluctuator model}

Several attempts were made to study the role of non-Gaussian and
non-Markovian nature of the $1/f$-noise for various examples of
coherent quantum transport such as resonant
tunneling~\cite{GaChao1}, ballistic transport through a quantum
point contact~\cite{Hessling}, Josephson effect~\cite{GG1},
Andreev interferometer~\cite{Lundin}.  In connection to qubits, a
similar model has been recently studied by Paladino \emph{et
  al.}~\cite{Paladino}. In this paper dynamical charged traps were
considered as two-level systems (TLS) with exponentially-broad
distribution of hopping rates, and the ``free induction signal'' of a
qubit was numerically analyzed for a narrow distribution of the
coupling constants, $v_i$, between the traps and the qubit.
Quantum aspects of non-Markovian kinetics were addressed in
Ref.~\onlinecite{Loss}.

The aim of the present work is to revisit this problem. In the
following we will consider a  \emph{spin-fluctuator} model,
similar to that considered in Ref.~\onlinecite{Paladino}, which
takes into  account both nonequilibrium and non-Markovian effects.
Analysis of this model  shows that nonequilibrium
effects are important. In particular, we will address the role of the
distribution of coupling constants $v_i$ between the \fs and the
qubit. The distribution of $v_i$ is probably broad for most possible
devices and realistic situations because the \fs are located at
different distances from the qubit. We will show that a broad
distribution of the coupling constants leads to significant
modifications of the decoherence dynamics.

The broad distribution of the coupling constants,
makes the model, which we will consider, essentially similar to the
conventional models of the \emph{spectral diffusion} in glasses.
The concept of spectral diffusion was introduced by Klauder and
Anderson~\cite{KlauderAnderson} as early as in 1962 for the problem of
spin resonance. They considered spins resonant to the external
microwave field (spins $A$) which generate echo signals, and
surrounding non-resonant spins (spins $B$). Due to interaction between
spins $A$ and $B$, stochastic flip-flops of spins $B$ lead to a random
walk of $A$-spins frequencies. This random walk is referred to as the
\emph{spectral diffusion}.

Black and Halperin~\cite{BlackHalperin} applied the concept of
spectral diffusion to low-temperature physics of glasses. They
used ideas of Ref.~\onlinecite{KlauderAnderson} to consider phonon
echo and saturation of sound attenuation by two-level
systems~\cite{AHVP} in glasses. Important generalizations of these
results were made by Hu and Walker~\cite{HuWalker}, Mainard
\emph{et al.}~\cite{Mainard} and Laikhtman~\cite{Laikhtman}.

\begin{figure}[t]
\centerline{
\includegraphics[width=6cm]{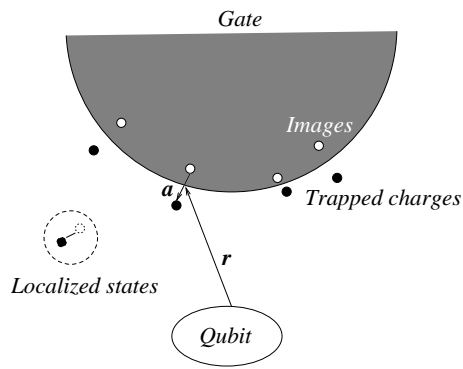}
}
\caption{Schematic sketch of the distribution of charged traps. They
  are located near the gate's surface and produce oppositely charged
  images.\label{fig1}}
\end{figure}

Following these ideas, we assume that the qubit is a two-level system
(TLS) surrounded by
so-called fluctuators, which also
are systems with two (meta)\-sta\-ble states. One can imagine
several realizations of two-level fluctuators. Consider, e.~g.,
structural dynamic defects, which usually accompany really
quenched disorder. These defects, if not charged, behave as
elastic dipoles, i.~e., they interact with the qubit via
deformational potential. The interaction strength in this case
depends on the distance, $r$, between the \f and the qubit as
$r^{-3}$. Consequently, the distribution of the coupling constants
is $\cP (v) \propto v^{-2}$.
 Charge traps near the gates also produce dipole electric fields, see
 Fig.~\ref{fig1}.
In this case $v=e^2(\ba\cdot \br)/r^3 \propto r^{-2}$. Assuming
that $a \ll r$ (otherwise charge would tunnel directly to the
qubit and the device will not work) and integrating the \fs
contributions along the 2D gate surface we again obtain $\cP(v)
\propto v^{-2}$. It is exactly the distribution of the coupling
constants in glasses, where two-level systems (TLS) interact via
dipole-dipole interaction~\cite{BlackHalperin}.

It is crucial that due to their interaction with the environment
the \fs \emph{switch} between their states. This switching makes
the fields acting upon the qubit \emph{time-dependent}. The
decoherence is caused by the time-dependence of the random field -
a static field can only renormalize the qubit's interlevel
spacing. The dynamics of the \fs is relaxational, so they are
rather ``relaxators'' than oscillators. The decoherence is thus
determined by the \fs relaxation rates, which should be compared
to the measurement time. At low temperatures the \fs are frozen in
the ground states, and their dynamics is slow. Therefore, the
fluctuators-induced decoherence significantly decreases with the
temperature.

The paper is organized as follows. In Sec.~\ref{Model} we
reformulate the model for spectral diffusion in glasses for the
case of a qubit and consider a simplified version of the theory.
This approach is, strictly speaking, applicable only to the \fs
with the small interlevel spacings, $E\ll T$, and the numerical
factors that follow from this approximation should not be trusted.
However the resulting qualitative physical picture is believed to
be correct, because the \fs with $E\gtrsim T$ are frozen in their
ground states, they thus do not fluctuate and do not contribute to
the decoherence.  The problem of a single neighboring \f is
considered in Sec.~\ref{Single} followed by the discussion of the
averaging over the ensemble of \fs in Sec.~\ref{Many}. In
Sec.~\ref{Flicker} we show that different \fs are responsible for
the decoherence and for the flicker noise. Consequently, the
decoherence cannot in general be expressed through the pair
correlation function of random fields acting upon the qubit.

\subsection{Detailed description}\label{Model}

A qubit coupled to the environment will be modeled by the Hamiltonian
\begin{equation}
  \label{eq:001}
  \tilde{\cH}=\tilde{\cH}_q +\tilde{\cH}_\text{man}+\tilde{\cH}_{qF}
  +\tilde{\cH}_{F}\, ,
\end{equation}
where $\tilde{\cH}_q$ and $\tilde{\cH}_F$ describe the qubit and
the fluctuators separately. A completely isolated qubit is just a
system that can be in one of two states and is characterized by
the energies of these states and the tunneling probability between
them. $\tilde{\cH}_q$ can thus be written as the Hamiltonian of
the qubit pseudospin in a static "magnetic field" $\bB
=\left\{B_x,B_z\right\}$, where $B_z$ characterizes the splitting
of the energies of the two states, and $B_x$ is responsible for
the tunneling. (Parallel shift of the qubit energies is, of course,
irrelevant). One can diagonalize such a Hamiltonian, by simply
choosing the direction of the new $z$-axis to be parallel to $\bB$
and write the rotated Hamiltonian $\cH_q$ as
\begin{equation}
  \label{eq:002}
  \cH_q =-\frac{1}{2} B\sigma_z \, .
   \end{equation}
 $\tilde{\cH}_F$ in turn can be diagonalized and  split into three parts
\begin{equation}
  \label{eq:003}
  \cH_F= \cH_{F}^{(0)} +\cH_{F\!\!-\text{env}}+\cH_{\text{env}}  \, .
\end{equation}
We model \fs by two-level \emph{tunneling systems}, which
Hamiltonians can also be diagonalized:
\begin{equation}
  \label{eq:003a}
  \cH_{F}^{(0)}=\frac{1}{2}\sum_i E_i \tau_z^{i}
\end{equation}
where the Pauli matrices $\bm{\tau}^{(i)}$ correspond to $i$-th
fluctuator. The spacing between the two levels, $E_i$ is formed by
the diagonal splitting, $\Delta_i$, and the tunneling overlap
integral, $\Lambda_i$
\begin{equation}
  \label{eq:thetai}
  E_i=\sqrt{\Delta_i^2 + \Lambda_i^2}\equiv
  \Lambda_i/\sin \theta_i \, .
\end{equation}

To account for the flip-flops of the \fs one needs to include
environment. To be specific we model the environment by a bosonic
bath. This applies not only to phonons, but also to electron-hole
pairs in conducting part of the system~\cite{Black}. Neglecting
the interactions of the \fs with the environment that do not cause
the flip-flops we specify the environment-related parts of the
Hamiltonian as
\begin{eqnarray}
\cH_{\text{env}}&=&\sum_\mu \omega_\mu  \left(\hat{b}_\mu^\dag
  \hat{b}_\mu +1/2 \right)\, ,  \label{eq:003b}\\
 \cH_{F\!\!-\text{env}}&=&\sum_i \tau_{x}^{(i)}
  \sum_{\mu} 
  C_{i,\mu} \left(\hat{b}_{\mu} + \hat{b}_\mu^\dag \right) \label{eq:003c}
\end{eqnarray}
 where $\mu$ represents the boson quantum numbers (wave vector,
etc.), and $C_{i,\mu}$ are the constants of the coupling between
the \fs and the bosons.

It is crucial to specify the interaction, ${\cH}_{qF}$, between
the qubit and the fluctuators. Following
Ref.~\onlinecite{BlackHalperin}, we assume that
\begin{equation}
  \label{eq:004b}
 \cH_{qF}=\sum_i v_i \,
 \sigma_z\tau_z^i\, , \quad  v_i=g(r_i)A(\bn_i)\cos \theta_i
\, .
\end{equation}
Here $\theta_i$ is determined by Eq.~(\ref{eq:thetai}), $\bn_i$ is
the direction of elastic or electric dipole moment of $i$th
fluctuator, and $r_i$ is the distance between the qubit  and $i$th
fluctuator. The functions $A(\bn_i)$ and $g(r_i)$ are not universal.

Interaction between the \fs and the environment manifests itself
through time-dependent random fields $\xi_i (t)$ applied to the
qubit. These low frequency, $\omega \ll T$, fields can thus be
treated \emph{classically}. Accordingly, one can substitute
$\cH_{qF}$ by the interaction of the qubit pseudospin with a
random, time-dependent magnetic field, $\cX_1(t)$, which is formed
by independent contributions of surrounding fluctuators:
\begin{equation}  \label{tp4}
\cH_{qF}=\cX_1(t)\, \sigma_z\,  , \quad
  \cX_1(t)=\sum_i v_i \xi_i(t) \, ,
\end{equation}
 so that $v_i$, Eq.~(\ref{eq:004b}) determines
 the coupling strength of the $i$-th \f
with the qubit, while the random function $\xi_i(t)$ characterizes
the state of this \f at each moment of time. Below we assume that
\f switching itself is an abrupt process that takes negligible
time, and thus at any given $t$ either $\xi_i(t)=0$ or
$\xi_i(t)=1$.

Note that we directed $\cX_1(t)$ along $z$-axis, i.e., neglected
possible transitions between the qubit eigenstates induced by the
random fields. This can be justified by the low frequency of these
field. (From the practical point of view a qubit that switches
frequently by the external noise does not make a decent device.)

To manipulate the qubit one should be able to apply ac ``magnetic
field'', $\bF (t)$. In general this field not only causes the
transitions between the qubit eigenstates ($F_x$), but also
modulates the qubit level spacing $B$ in time ($F_z$). The latter
effect is parasitic and should be reduced. Here we simply neglect
this modulation and assume that $\bF$ is parallel to the $x$-axis
($F_x = F$). Accordingly
\begin{equation}
  \label{eq:004a}
  \cH_\text{man}=(1/2)F(t)\cdot\sigma_x\, .
 \end{equation}
For the manipulation to be resonant, the frequency $\Omega$ of the
external field $F$ should be close to $B$.

Combining Eqs.~(\ref{eq:002}), (\ref{eq:003a}), (\ref{tp4}), and
(\ref{eq:004a}) we substitute the initial Hamiltonian
(\ref{eq:001}) by:
\begin{equation}
  \label{eq:005a}
  \cH =\frac{1}{2} \left[E_0+\cX(t)\right] \sigma_z
  +\frac{1}{2} F (t)\sigma_x
  +\frac{1}{2} \sum_i E_i \tau_z^{i}\, .
\end{equation}
Here $E_0$ determines the original ($t=0$) value of the
eigenfrequency of the qubit, while $\cX(t)$ is the random
modulation of this eigenfrequency caused by the flips of the
fluctuators:
\begin{equation}
E_0 = B+\cX_1(0)\, , \quad
\cX(t) =\cX_1(t) - \cX_1(0) \, .
\label{eq:005b}
  \end{equation}

We have not explicitly included into the Hamiltonian (\ref{eq:005a}) the
interaction (\ref{eq:003b}) of the \fs with phonons or electrons,
which causes flips of the fluctuator. We account for these flips by
introducing finite transition rates between the fluctuator's states.
The transition rates can be
calculated in the second order of the perturbation theory for the
fluctuator-phonon/electron interaction~\cite{Jackle,Black}. As a
result, the
ratio, $\gamma^{+}/\gamma^{-}$, of the rates for the upper,
$\gamma^{+}$, and the lower, $\gamma^{-}$, states in the
equilibrium is determined by the energy spacing between these two
states: $\gamma_i^{+}/\gamma_i^{-} = \exp\left(E_i/T\right)$.
Accordingly, these rates can be parameterized as
\begin{equation}
  \label{eq:006}
  \gamma_i^{\pm} = (1/2)\gamma_i\left(T,E_i\right)\,
  \left[1\pm \tanh(E_i/T)\right]\,
  \sin^2\theta_i \,
\end{equation}
where $\theta_i$ is given by Eq.~(\ref{eq:thetai}).

The dependence of $\gamma_0 (E_i)$ on $E_i$ is determined by the
concrete relaxation mechanism: for phonons~\cite{Jackle} $\gamma_0
(E_i) \propto E_i^3$, while for electrons~\cite{Black} $\gamma_0
(E_i) \propto E_i$. Note that the average value of $\xi_i(t)$
depends only on $E_i$ and $T$:
\begin{equation}
  \label{eq:n0}
  \langle \xi(t)\rangle_{t \to \infty} =
  \gamma^{-}/(\gamma^{+}+\gamma^{-})
  =[1+\exp\left(E_i/T\right)]^{-1}\, .
\end{equation}
There can also be a direct interaction of the qubit with the
bosonic bath. One can introduce the transition rate $\gamma_q(T)$
due to this interaction in the same way as $\gamma_i (T)$.

Below we will often use a simplified version of the theory
assuming that the only \fs that contribute to dephasing are those
with $E_i \ll T$. In this approximation
\begin{equation}
  \label{eq:007}
  \gamma_i^{\pm}= (1/2)
  \gamma_0(T) \sin^2 \theta_i \, ,
\end{equation}
$\gamma_0$ is thus
the \emph{maximal} fluctuator relaxation rate at a given
temperature $T$. This assumption significantly simplifies the
formulas and produces, as one can show~\cite{Laikhtman,GPM},
correct order-of-magnitude results.

The model formulated above differs from the spin-boson models by
statistics of the random force $\cX(t)$. It allows one to describe
the qubit decoherence in a simple way without loosing track of the
essential physical picture.

\subsection{Qubit manipulations}

We parameterize the qubit's density matrix as
\begin{equation}
  \label{eq:012}
  \hat{\rho}=\left(\begin{array}{cc}
n&-if\, e^{i\Omega t}\\
if^*\, e^{-i\Omega  t}& 1-n \end{array}
 \right)\, ,
\end{equation}
and commute it with the Hamiltonian, Eq.~(\ref{eq:005a}), using
the resonance approximation. (The frequency, $\Omega$ of the
applied field $F$ is assumed to be close to the qubit
eigenfrequency, $E_0$.) Including also the inherent qubit relaxation
($\gamma_q$) we obtain the following equations of motion:
\begin{eqnarray}
  \label{eq:013}
&& \frac{\partial n}{\partial t} = -2\gamma_q(n-n_0) -F\Re f\, , \\
&& \frac{\partial f}{\partial t}=i\left[ E_0 +\cX(t)- \Omega
\right] f
 -\gamma_q f
 +F\left( n-\frac{1}{2}\right).
\end{eqnarray}
These equations have been obtained in Refs.~\onlinecite{GGP1,GPM2}
for the problem of spectral diffusion in glasses, $F$ in
Eq.~(\ref{eq:005a}) playing the role of the Rabi frequency of the
resonant pair. 

The equation set (\ref{eq:013}) belongs to the class of
\emph{stochastic differential equations}.  In the following we use
the methods~\cite{Brissaud1,Kampen,Klyatskin,Brissaud2,Loginov,Anderson2,Kubo}
developed for these equations to study the qubit response to
various types of the manipulation.

Currently the experimentally observable signal is an accumulated
result of numerous repetitions of the same sequence of inputs
(e.~g., pulses of the external field). To be compared with such
measurements, solutions of Eq.~(\ref{eq:012}) should be averaged
over realizations of the stochastic dynamics of the fluctuators.
 Provided that the time intervals between the
sequences of inputs are much longer than the single-shot measuring time  
we can separate the averaging over the initial
states of the fluctuators, $\xi_i (0)$, from the averaging over
their stochastic dynamics, $\cX(t)$.

In the absence of the external ac field $n(t)$ should approach its
equilibrium value given by the Fermi function, $n_0(E_0)=\left[1+\exp(E_0/T)\right]^{-1}$,
while the off-diagonal matrix element of the density matrix should
tend to zero. If the external ac field is switched off at $t=0$
then the solution of Eq.~(\ref{eq:012}) has the form
\begin{eqnarray}
  n(t)&=&n_0(E_0)+[n(0)-n_0(E_0)]e^{-2\gamma_q t}\, ,  \label{eq:013a} \\
f(t)&=& f(0)\, e^{ -\gamma_q t +i(E_0 - \Omega)t+i\int_0^t
\cX(t')\, dt'} \, . \label{eq:013b}
\end{eqnarray}
We need to average this signal, known as the free induction signal
over both $E_0$ and $\cX(t)$. If the time $t$ after the pulse is
short enough, then the \fs remain in their original states and
$\cX(t)$ can be neglected. At $t=0$ the system of the \fs is
supposed to be in the equilibrium, and thus probability for $\xi_i
=1$ is $n_0 (E_i)$. Substituting $E_0$ from Eq.~(\ref{eq:005b}) we
find that $\exp [i(E_0-\Omega)t]$ averaged over the initial
realization equals to
\begin{equation} \label{eq:tmp01}
\left \langle e^{i(E_0-\Omega)t} \right \rangle_{\! \xi}
  =e^{i(B-\Omega) t} \prod_j\left[1-n_0 (E_j)+n_0 (E_j)e^{iv_j
  t}\right]\, .
\end{equation}

 It follows from Eq.~(\ref{eq:tmp01}) that the observed free
induction signal involves the oscillations with frequencies that
differ from $B - \Omega$ by various combinations of $v_j$. As a
result, in the presence of a large number of the fluctuators the
free inductance signal decays in time even when $\cX(t)=0$, i.~e.
when the \fs do not switch during one experimental run. However,
this decay has little to do with the decoherence due to
irreversible processes.

Much  more informative for studies of genuine decoherence are echo
experiments, when the system is subject to two (or three) short ac
pulses with different durations $\tau_1$ and $\tau_2$, the time
interval between them being $\tau_{12}$ (or $\tau_{12}$ and
$\tau_{13}$, respectively). Considering echo, we assume that the
external pulses are short enough for both relaxation and spectral
diffusion during each of the pulses to be neglected. The echo
decay is known to be proportional to the "phase-memory
functional"~\cite{Mims}
\begin{equation}
  \label{eq:016}
  \Psi [\beta(t'),t]= \left \langle \exp \left(i \int_0^t \beta(t') \cX(t')\, dt' \right)
  \right \rangle_{\xi_i}   \, .
\end{equation}
Here for 2-pulse echo $t=2\tau_{12}$ and
\begin{equation}
  \label{eq:015}
  \beta (t')=\left\{\begin{array}{rcc}
0&\text{for}&t'\le 0\, ,\\
1&\text{for}&0<t' \le \tau_{12}\, ,\\
-1&\text{for}&\tau_{12}<t'\, .
\end{array}\right.
\end{equation}
In the case of the 3-pulse echo one would put $t=\tau_{12}
+\tau_{13}$ and
\begin{equation}
  \label{eq:015a}
  \beta (t')=\left\{\begin{array}{rcc}
0&\text{for}&t'\le 0\, ,\\
1&\text{for}&0<t' \le \tau_{12}\, ,\\
0&\text{for}&\tau_{12} <t' \le \tau_{13}\, ,\\
-1&\text{for}&\tau_{13}<t'\, .
\end{array}\right.
\end{equation}
The functional (\ref{eq:016}) can be used to describe the
decoherence of the free induction signal, substituting $\beta
(t')=\Theta (t')$. In this case, however, one should understand
$\cX(t')$ as $\sum_i v_i\,\xi_i (t')$ rather than use
Eq.~(\ref{eq:005b}). For the echo
experiments $t$ in Eq.~(\ref{eq:016}) is chosen in such a way that
the integral of $\beta (t')$ from zero to $t$ vanishes. As a
result, the time-independent part of $\cX$, i.~e., dispersion of
$\xi_i (0)$ becomes irrelevant. Below we evaluate the phase-memory
functional (\ref{eq:016}) for the free induction as well as for
schemes of the measurement.

\section{Results for a single fluctuator} \label{Single}

\subsection{Random telegraph noise}
The process, which is described by a random function, $\xi (t)$,
that acquires only two values: either $\xi =0$ or $\xi =1$ is
known as \emph{random telegraph process}. In this section we
assume that $E_i \gg T$ and thus the two states of each fluctuator
are statistically equivalent, i.e., the time-average value of $\xi
(t)$ equals to $\langle \xi (t)\rangle_{t \to \infty} =1/2$.

To evaluate the memory functional (\ref{eq:016}) we first
introduce auxiliary random telegraph processes defined as
\begin{equation} \label{tp1}
z_\pm(t)=\pm (-1)^{n(0,t)}\, ,
\end{equation}
where $n(t_1,t_2)$ is a random sequence of integers describing
number of 'flips" during the period $(t_1,t_2)$, so that
$n(t,t)=0$. The fact that $\ z_\pm^2(t)=1$ substantially
simplifies the calculations. The  'flips"  of a given \f induced
by its interaction with the bosonic bath should not be correlated
with each other. Accordingly, $n(t_1,t_2)$ obeys the Poisson
distribution, i.~e., the probability, $\cP_{n}(t_1,t_2))$, that
$n(t_1,t_2)=n$ equals to
\begin{equation} \label{Pd1}
\cP_{n}(t_1,t_2))=\frac{\av{n(t_1,t_2)}^n}{n!}\,
e^{-\av{n(t_1,t_2)}} \, ,\quad \av{n(t_1,t_2)}= \gamma\,
|t_1-t_2|\, ,
\end{equation}
where $\gamma$ has a meaning of the average frequency of "flips".
{}From Eqs.~(\ref{Pd1}) and (\ref{tp1}) it follows that
\begin{eqnarray*}
\langle z_\pm(t)\rangle&=& \sum_{n=0}^\infty (-1)^{n}\cP_{n} =
\pm  e^{- 2\gamma|t|} \, ,
\\
\langle z_\pm(t_1) z_\pm(t_2)\rangle &=&  \langle
(-1)^{n(t_2,t_1)}\rangle =  e^{-2 \gamma (t_1-t_2)}, \ t_1 \ge
t_2\, .
\end{eqnarray*}

It is convenient to describe different measurement schemes by
making use of the generating functionals
\begin{equation}
  \label{eq:genf01}
  \psi_\pm[\beta,t]=\left \langle \exp
\left[\frac{iv}{2}\int_0^t \beta(t')z_\pm(t') dt'\right]
  \right \rangle_{\! \! z_\pm(t)}\, ,
\end{equation}
where $\beta (t')$ is the same function as in Eqs.~(\ref{eq:016}),
(\ref{eq:015}), and  (\ref{eq:015a}), while the constant $v$ will
later play the role of the qubit-fluctuator coupling constant. To
evaluate the functionals (\ref{eq:genf01}) consider a set of the
correlation functions
$$M_n(t_1, t_2, \ldots,t_n) \equiv \langle
z_\pm(t_1)z_\pm(t_2)\cdots z_\pm(t_n) \rangle  \, , \quad t_1 \ge
t_2\ge  \ldots \ge t_n\, .$$ It is convenient to use a recursive
formula
\begin{equation} \label{eq:recursion}
M_n(t_1,t_2,\ldots,t_n)=e^{-2
\gamma(t_1-t_2)}M_{n-2}(t_3,\ldots,t_n) \, .
\end{equation}
which follows directly from Eqs.~(\ref{Pd1}) and (\ref{tp1}).
Combining Eq.~(\ref{eq:recursion}) with the Taylor expansion of
Eq.~(\ref{eq:genf01}) we obtain an exact integral equation for
$\psi_\pm[\beta,t]$:
\begin{eqnarray}
&&\psi_\pm (\beta,t)= 1\pm i(v/2)\int_0^t dt_1 e^{-2 \gamma
t_1}\beta(t_1) \nonumber \\ && \quad -(v^2/4)\int_0^t \! dt_1
\int_0^t \! dt_2 \, e^{-2 \gamma (t_1
-t_2)}\beta(t_1)\beta(t_2)\psi [\beta,t_2] \, . \label{eq:02}
\end{eqnarray}
One can evaluate second time-derivative of both sides of
Eq.~(\ref{eq:02}) and transform this integral equation into a
second order differential equation~\cite{Klyatskin}
\begin{equation} \label{tp2a}
\frac{d^2 \psi_{\pm}}{dt^2}+\left[2 \gamma-\frac{d\ln \beta(t)}{dt}\right] \frac{d\, \psi_{\pm}}{dt} +
\frac{v^2}{4} \psi_{\pm}=0
\end{equation}
 with initial conditions
\begin{equation} \label{tp3a}
\psi_{\pm}(0)=1\, , \quad \left.\frac{d\,
    \psi_{\pm}}{dt}\right|_{t=0}=\pm \frac{iv}{2}\beta(t=-0)\, .
\end{equation}

\subsection{Generating functional}

In the limit $E_i \ll T$ the random functions $\xi_i(t)$ can be
expressed through $z_{+}(t)$ or by $z_{-}(t)$ with equal
probabilities. Using Eq.~(\ref{eq:005b}) we thus can rewrite the
memory functional (\ref{eq:016}) in terms of the functionals
$\psi_{\pm}$, Eq.~(\ref{eq:genf01}):
\begin{equation}
  \label{eq:genf03}
 \psi [\beta ,t]=\frac{1}{2}\sum_{\pm} e^{\mp i(v/2)\int_0^t \beta
 (t')\, dt'}\psi_\pm [\beta (t),t]  \, .
\end{equation}
{}From Eqs.~(\ref{eq:genf03},\ref{tp2a}) follows the differential
equation for
 $ \psi [\beta ,t]$:
\begin{equation}
  \label{eq:dur01}
 \frac{d^2 \psi}{dt^2}+\left[2 \gamma-\frac{d\ln \beta(t)}{dt}-iv
 \beta (t) \right]
 \frac{d\, \psi_{\pm}}{dt} -iv \gamma \psi=0
\end{equation}
with initial conditions
\begin{equation}
  \label{eq:dur02}
 \psi(0)=1\, , \quad \left.\frac{d\, \psi}{dt}\right|_{t=0}=i\beta (-0)
 \frac{v}{2}\, .
\end{equation}
Below we use this equation to analyze decay of the free induction
and echo signals. Note that Eq. (\ref{eq:dur01}) is the $E_i/T \to
0$ limit of equation derived in Ref.~\onlinecite{Klyatskin} for
arbitrary $E_i/T$.

For the free induction signal $\beta(t>0)=1$ and  $\beta(-0)=0$.
 For this $\beta$-function Eq.~(\ref{eq:dur01}) with initial
conditions (\ref{eq:dur02}) yields the following phase memory
functional
\begin{equation} \label{eq:104c}
  \psi_{pm} (t)=\frac{e^{-\gamma t}}{4\mu}\sum_{q=\pm1}\sum_{p=\pm
  1} p\, \left(1-q\frac{iv }{2\gamma}+p\mu \right)e^{-(iqv/2
  +p\gamma \mu) t}\, .
\end{equation}
where $\mu=1-\left(v/2\gamma\right)^2$. This solution is the
$E_i/T \to 0$ limit of the result obtained in
Ref.~\onlinecite{Laikhtman}. At short times, $\gamma t \ll 1$,
Eq.~(\ref{eq:104c}) can be approximated as
\begin{equation}
  \label{eq:104b}
 \psi_{pm} (t) =1-\gamma \left( t-\frac{\sin vt}{v}\right) \, .
\end{equation}

However $\psi_{pm}$ given by Eq.~(\ref{eq:104c}) does not describe
the free induction decay, $\exp \left[i v\int_0^t \xi(t')\,
dt'\right]$, even within our simplified model. We have to consider
\begin{equation}
  \label{eq:psifi}
  \psi_{fi}(t)=\exp \left(\frac{ivt}{2}\right)\frac{\psi_++\psi_-}{2}
\end{equation}
rather than $\psi_{pm}(t)$, because the latter neglects the
dispersion of $E_0$, (\ref{eq:005b}), which is due to the term
$v\xi (0)$. {}From Eqs.~(\ref{tp2a}) and (\ref{tp3a}) it follows that
\begin{equation}
  \label{eq:frind01}
  \psi_{fi}(t)=e^{(iv/2 -\gamma)t}
  \left(\mu^{-1}\cosh \gamma \mu t+\sinh \gamma \mu t
  \right)\, .
\end{equation}
\begin{figure}[ht]
\sidebyside {\includegraphics[width=.3\textwidth]{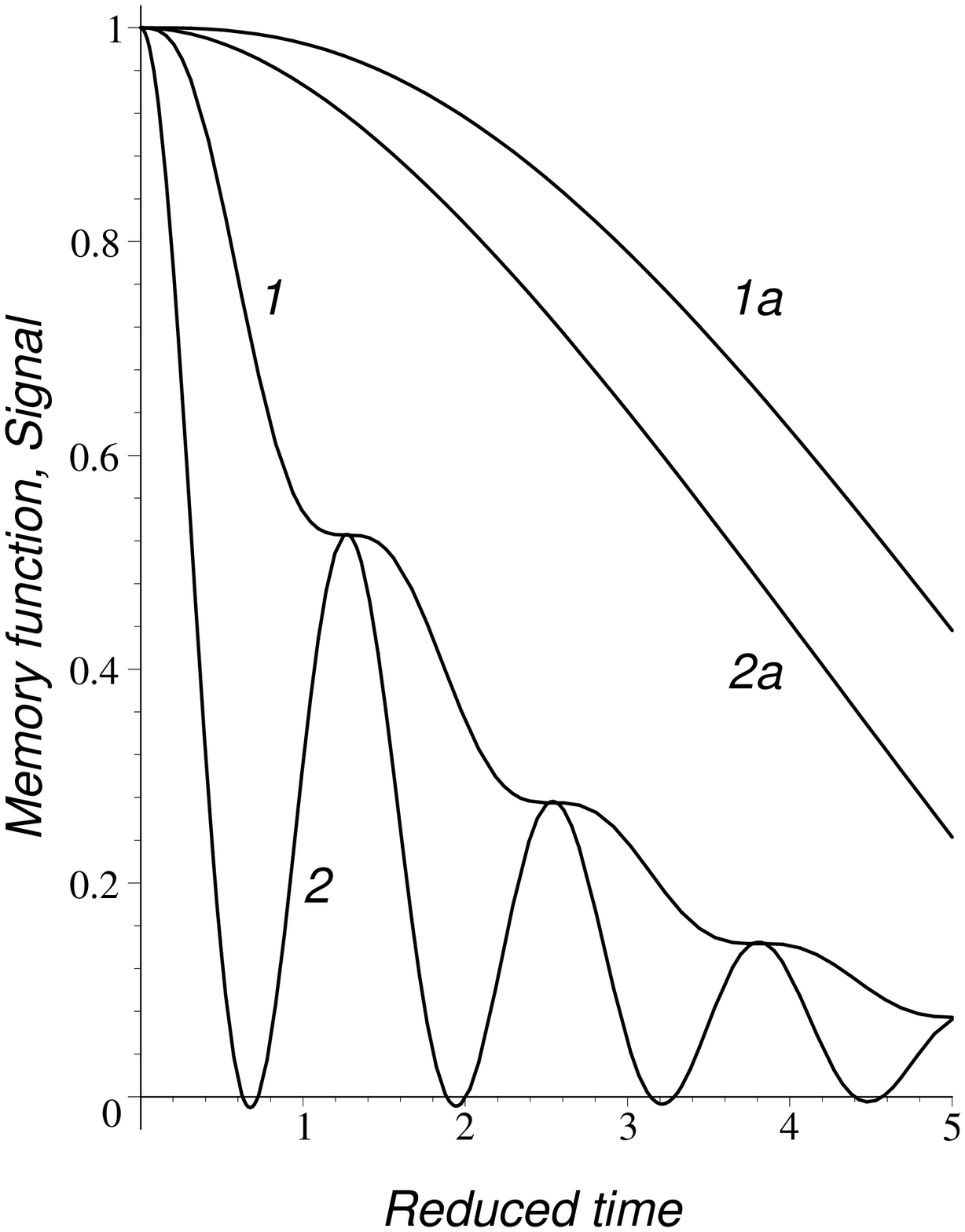}
\caption{Comparison of the phase memory function (1,1a) with the
  free induction signal (2,2a).
   $v/2\gamma=5$ (1,2), and  0.5  (2,2a). Time is measured
   in units of $(2\gamma^{-1})$.
  \label{fig:comp1}}}
{\includegraphics[width=.3\textwidth]{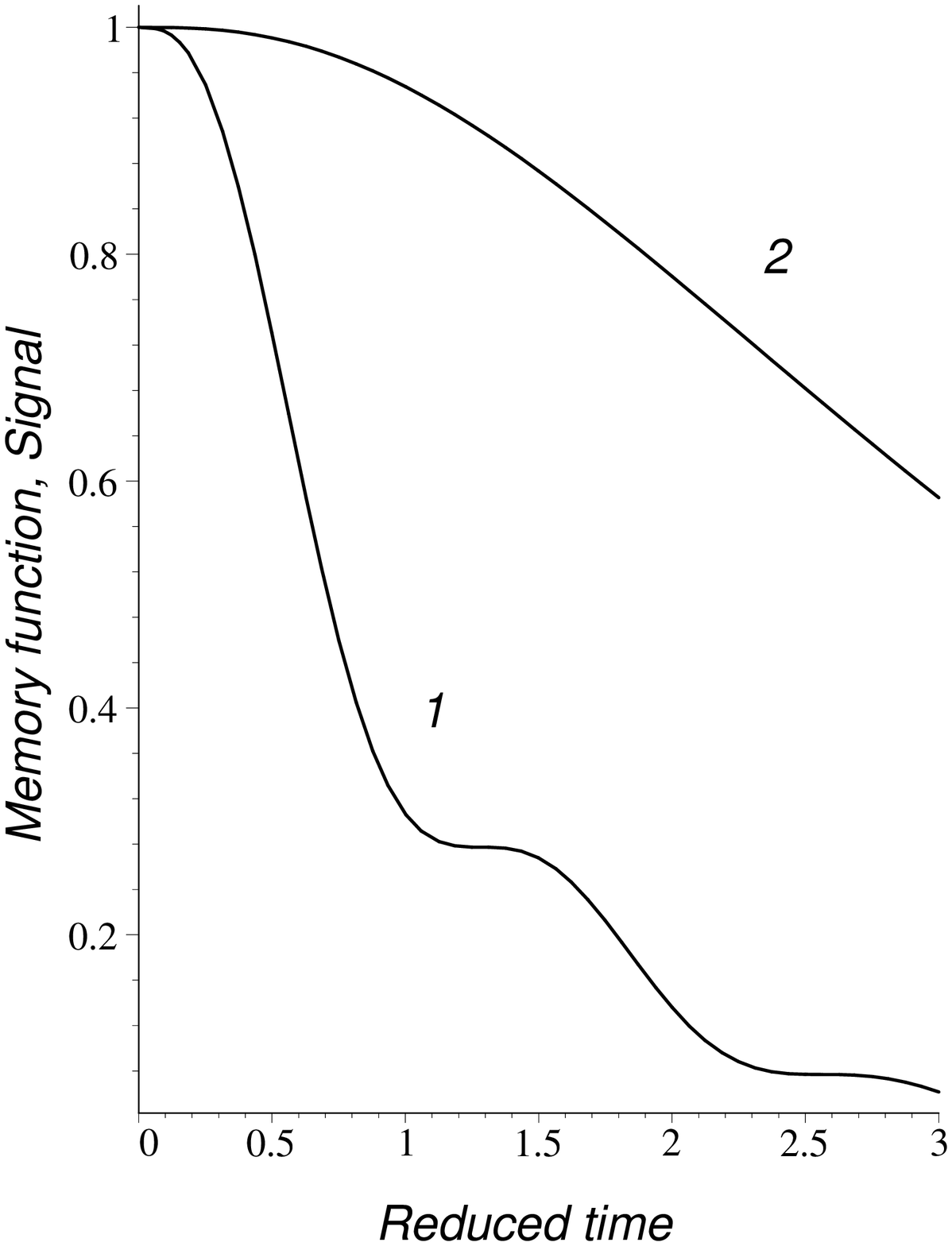} \caption{Two-pulse
echo signal for $v/2\gamma=5$ (1) and 0.8
  (2). Time is measured
   in units of $(2\gamma^{-1})$. \label{fig:04}}}
\end{figure}
This expression is the $E_i/T \to 0$ limit of the result obtained
in Ref.~\onlinecite{Paladino}. Comparison of the free induction
decay ~(\ref{eq:frind01}) with decay of the phase memory is
presented in Fig.~\ref{fig:comp1}. This difference is especially
important for the case of $1/f$-noise, when many fluctuators
contribute.

The calculation of the echo decay can be done in a similar
way. The results for 2- and 3-pulse echo are, respectively (cf. with
Ref.~\onlinecite{Laikhtman}):
 \begin{eqnarray}
  &&  \psi_{e2}(2\tau_{12})=\frac{e^{-2\gamma\tau_{12}
   }}{2|\mu|^2}
\sum_\pm
\left [ (1+\mu_2^2) (1\pm\mu_1)e^{\pm 2\mu_1\gamma
   \tau_{12}}
 \right. \nonumber \\ && \qquad \left.
-(1-\mu_1^2)(1\mp i \mu_2)e^{\mp 2i \mu_2\gamma \tau_{12}}
 \right]
\, ,  \label{eq:04} \\
&&\psi_{e3}(\tau_{12}+\tau_{13})=\psi_{e2}(2\tau_{12})
 +\frac{1}{2}
   \left(\frac{v}{2|\mu|\gamma}\right)^2 \left(\cos 2\mu_2
   \tau_{12}
\right. \nonumber \\ && \qquad \left.
-\cos 2\mu_1
   \tau_{12}\right)
\left(1-e^{-2\gamma(\tau_{13}-\tau_{12})}\right)e^{-2\gamma
   \tau_{12}}\, .  \label{eq:04c}
 \end{eqnarray}
Here $\tau_{13}$ is the time between first and third pulse,
$\mu_1+i \mu_2= \sqrt{1-(v/2\gamma)^2}$. The function
$\psi_{e2}(2\tau_{12})$ is shown in Fig.~\ref{fig:04}. Note that
at $v > 2\gamma$ the time dependence of the echo signal shows
steps similar to what was experimentally observed for the charge
echo~\cite{Nak}. The expressions for the echo signal have a simple
form at $v \gg \gamma$. In particular, the expression for the
two-pulse echo acquires the form
$$\psi_{e2}(2\tau_{12})=e^{-2\gamma\tau_{12}}\left([1+(2\gamma/v)\, \sin v
\tau_{12}\right]\, .$$ Consequently, the plateaus occur at $v
\tau_{12}=k \pi/2 -\arctan(2\gamma/v)$ and their heights
exponentially decay with the number $k$.

\section{Summation over many fluctuators} \label{Many}

To average over a set of the fluctuators we assume that dynamics
of different fluctuators are not correlated, i.e., $\langle
\xi_i(t) \xi_j(t') = 0$, unless $i=j$. Under this condition the
generating functional is a product of the partial functionals,
$\psi^{(i)}$. Hence, the generating functional can be
presented as
\begin{equation}
  \label{eq:008}
\Psi[\beta(t),t] = \prod_i \psi^{(i)} (t)   =e^{\sum_i \ln
\psi^{(i)} (t) }\equiv
 e^{-\cK   (t)}\, .
\end{equation}
Since logarithm of the product is a self-averaged quantity, at
large number of fluctuators, $\cN \gg 1$,  one can replace the
sum $\sum_i \ln \psi^{(i)}$ by $\cN \langle \ln \psi
\rangle_{\!F}$, where $\langle \cdots \rangle_{\!F}$ denotes
average over the fluctuators interlevel spacings $E_i$, their
interaction strength, $v_i$, and tunneling parameters, $\sin
\theta_i$. Furthermore, for $\cN \gg 1$ one can employ the Holtsmark
 procedure~\cite{Chandrasekhar}, i.~e., to
 replace $\langle \ln \psi
\rangle_{\!F}$ by $\langle \psi -1 \rangle_{\!F}$, assuming that
each of $\psi^{(i)}$ is close to $1$. Thus, for large $\cN$ the
approximate expression for $\cK (t)$ is
\begin{equation}
  \label{Holtsmark}
\cK(t)\approx \sum_i \left(1-\psi^i(t)\right)=\cN \left \langle
  1-\psi (t) \right\rangle_{\!F}
\, .
\end{equation}

To evaluate $\cK(t)$ one has to specify the distributions of the
parameters $E_i$, $v_i$, and $\theta_i$ that characterize the
fluctuators.
 Taking into account only the fluctuators with
$E_i\lesssim T$ we can write the number of fluctuators per unit
volume as $P_0 T$. It is natural to assume that the density of
states, $P_0$ is a $T$-independent constant as it is in structural
glasses~\cite{AHVP}.

The conventional estimation of the distribution function of
relaxation rates is based on the fact that the tunneling splitting
$\Lambda$ depends exponentially on the distance in real space
between the positions of the two-state fluctuator (on the distance
between the charge trap and the gate), as well as on the height of
the barrier between the two states Assuming that parameters like
distances and barrier heights are distributed uniformly, one
concludes that $\Lambda$-distribution is $\propto \Lambda^{-1}$.
Since $\gamma$ is parameterized according to Eq.~(\ref{eq:007}),
 this implies $\cP(\theta) = 1/\sin \theta$. As a result, cf.
with Ref.~\onlinecite{AHVP},
\begin{equation}
  \label{eq:11}
  \cP(E,\theta)=P_0/\sin \theta \, .
\end{equation}

The coupling constants, $v_i$, determined by Eq.~(\ref{eq:004b}),
contain $\cos \theta_i$ and thus are statistically correlated with
$\theta_i$. It is convenient to introduce an uncorrelated random
coupling parameter, $u_i$ as
\begin{equation}
  \label{eq:coupling}
 u_i=g(r_i)A(\bn_i)\, , \ v_i=u_i\cos \theta_i\, .
\end{equation}

It is safe to assume that direction, $\bn_i$, of a fluctuator is
correlated neither with its distance from the qubit, $r_i$, nor
with the tunneling parameter $\theta_i$ and thus to replace
$A(\bn)$ by its angle average, $\langle |A(\bn)|\rangle_\bn$.  It
is likely that the coupling decays as power of the distance, $r$:
$g(r)\propto \bar{g}/r^b$. If the \fs are located near a
$d$-dimensional surface, then
 \begin{equation}
  \label{eq:06}
\cP (u,\theta)=\frac{{\eta}^{d/b} }{\sin\theta}\, u^{-d/b-1}\, , \
\eta=\frac{\bar{g}}{r_T^b}\, , \ r_T=\frac{a_d}{( P_0
    T)^{1/d}} \,  .
\end{equation}
Here $a_d$ is a dimensionless constant depending on the
dimensionality $d$ while $r_T$ is a typical distance between the
fluctuators with $E_i\lesssim T$. In the following we will for
simplicity assume that
\begin{equation}
\label{eq:mf} r_{\min} \ll r_T \ll r_{\max}\, ,
\end{equation}
 where $r_{\min}$ ($r_{\max}$) are distances between the qubit and the nearest
(most remote) fluctuator. Under this condition $\eta \propto
T^{b/d}$ is the typical constant of the qubit - fluctuator
coupling. As soon as the inequality (\ref{eq:mf}) is violated the
decoherence starts to depend explicitly on either $r_{\min}$ or
$r_{\max}$, i.~e., become sensitive to mesoscopic details of the
device. This case will be analyzed elsewhere. In the following we
assume that $d=b$, as it is for charged traps located near the
gate electrode, see Fig.~\ref{fig1}.

The dependences of the generating functional $\psi (\beta,t|u,
\gamma)$ on the coupling constants $u$ and transition rates
$\gamma$ of the fluctuators are determined by Eq.~(\ref{eq:016}).
Substituting equations (\ref{eq:007}) and (\ref{eq:coupling}) into
(\ref{eq:016}) and the result - into (\ref{Holtsmark}) one obtains
$\cK (t)$ in the form:
 \begin{equation}
   \cK (t)=\eta \int \frac{du}{u^2}\int_0^{\pi/2} \frac{d \theta}{\sin
   \theta}
   \left\{ \phantom{\frac{}{}}
   1-\psi\left[\beta,t|u\cos \theta, \gamma_0\sin^2 \theta \right]\right \}\, .
   \label{eq:07}
 \end{equation}
This expression allows one to calculate the dephasing rate in the
case of many surrounding \fs for various qubit manipulations.

To start with  consider the phase memory functional with
$\beta(t')=\Theta(t')$, i.~e., free induction signal will
$\xi_i(0)=0$. At small times, $t \ll \gamma_0^{-1}$, one can use
Eq.~(\ref{eq:104b}) for $\psi_{pm}(t)$. The integration over $u$
yields
$$\int_0^\infty \frac{du}{u^2}
[1-\psi_{pm}(t)]=\frac{\pi}{4}\, t^2\gamma_0\sin^2\theta \, .$$
Performing the following integral over $\theta$ one obtains
$\cK_{pm}\propto \eta \gamma_0 t^2$.

It is slightly trickier to estimate $\cK_{pm}(t)$ at large times,
$\gamma_0 t \gg 1$. One can show that the decoherence in this
limit is due to the fluctuators, which coupling with the qubit is
atypically weak: $u \sim t^{-1} \ll \gamma_0$. As a result, in
the leading in $1/(\gamma_0 t)$  approximation $\psi_{pm}(t) =\cos
u t/2$. This asymptotics can be interpreted in the following way.
At $ t \gg 1/\gamma_0$ a typical fluctuator had flipped many times
and its contribution to the qubit phase, which is proportional to
$$\left \langle \int_0^t [\xi(t')-\xi(0)] \, dt'\right \rangle \propto t\,  , $$
and does not depend on $\gamma$ and, hence on $\theta$. Therefore
the integral over $\theta$ in (\ref{eq:07}) diverges
logarithmically. The proper cut-off is determined by the condition
$\gamma t \approx 1$, i.~e., is the value of $\theta =
\theta_{min}(t)$, which allows approximately one flip during the
time $t$. Using Eq.~(\ref{eq:007}) we estimate $\theta_{min}(t)$
as $(\gamma_0 t)^{-1/2})$. This cut-off reflects the fact that the
\fs with too low tunneling rates do not change their states during
the measurement time $t$.

The estimate for $\cK_{pm}(t)$ can be then summarized as (cf. with
Ref.~\onlinecite{Laikhtman}),
\begin{equation}
  \cK_{pm}(t) \approx \eta \cdot\left\{
\begin{array}{lcl}
 \label{eq:09}
\gamma_0 t^2 & \text{for} & \gamma_0 t \ll 1 \, ;  \\
t \ln \gamma_0 t& \text{for} & \gamma_0 t \gg 1 \, .
\end{array}
\right.
\end{equation}
Now we can define  the dephasing time $\tau_\varphi$ by the
condition $\cK (\tau_\varphi)=1$. Using Eq.~(\ref{eq:09}), we get
\begin{equation}
  \label{eq:10}
  \tau_\varphi = \max \left \{\eta^{-1} \ln^{-1}(\gamma_0/\eta),(\eta
    \gamma_0)^{-1/2}\right \}\,.
\end{equation}
The  echo decay can be calculated in a similar way, cf. with
Ref.~\onlinecite{Laikhtman}:
\begin{eqnarray}
\cK_{e2}(2\tau_{12}) &\sim& \eta \tau \cdot \min\{1, \gamma_0 \tau_{12}\}\, , \\
\cK_{e3}(\tau_{12}+\tau_{13}) &\sim& \eta \tau_{12} \min\{1,  \ln
\tau_{13}/\tau_{12}\} \, .
\end{eqnarray}
The dephasing time for the two-pulse echo decay is then
\begin{equation}
  \label{eq:10a}
  \tau_\varphi =\max\left \{\eta^{-1},(\eta \gamma_0)^{-1/2}\right \}\,.
\end{equation}

Let us discuss the physical meaning of the
results~(\ref{eq:09})--(\ref{eq:10a}). If there is no flips of the fluctuators, then
the contribution of a given \f to
the total phase gain during the observation time is  $\xi(0)\int_{0}^t
\beta(t')\, dt'$. During the time interval $t \ll \gamma_0^{-1}$
each fluctuator can flip only once.
If it flips at time $t_1$ , the accumulated
relative phase is $\pm 2\int_{t_1}^t \beta(t')\, dt'$.  To obtain the total
phase gain one has to average over all possible moments of flips:
$$|\delta \varphi (t)| \sim  \gamma_0  \int_0^t\, dt_1 \left|\int_{t_1}^t
\beta (t')\, dt'\right|\, .$$
Since $v \propto r^{-3}$, nearest neighbors are important, and typical
value of $v$ is $\eta$. Thus we immediately obtain $\cK(t) \sim \eta
\gamma_0t^2$. It can be shown~\cite{BlackHalperin} that in this case
the random process is Markovian. Consequently in this case the
situation  can be characterized
by a pair conditional probability $K(E,t|E_0)$ to find the
spacing $E$ at time $t$ under condition that at $t=0$ it was
$E_0$. It has the Lorentzian form,
\begin{equation}
  \label{eq:12}
  K(E,t|E_0)=\frac{1}{\pi}\, \frac{\Gamma
  (t)}{(E-E_0)^2 + \Gamma^2 ( t)}\, ,
\end{equation}
where $\Gamma(x) \sim \eta \gamma_0 t$. This time dependence can be
easily understood in the following way. The nearest region of $r$,
where \emph{at least one} \f flips during the time interval $t$ gives
the maximal contribution. The size of this region can be estimated
from the condition $P_0Tr^3\gamma_0 t \approx 1$ that yields $r^{-3}
\approx P_0T \gamma_0 t$. The corresponding change in the qubit's
interlevel spacing is then given by the interaction strength at this
distance, $\eta \gamma_0 t$.

During a time interval $t \gg \gamma_0^{-1}$ a substantial
contribution comes from the \fs with $\gamma^{-1}_0 \ll  \gamma^{-1}
\ll t$, which experience many
flip-flops. Since a \f having $\xi=1$ can flip only to the state with
$\xi=0$ the contributions of successive hops are not statistically
independent. The density of most important \fs is of the order
$P_0Tr^3\ln \gamma_0 t$, and the substantial region of $r$ is
determined by the relation $P_0Tr^3\ln \gamma_0 t \sim 1$. As a
result $\Gamma (t) \sim \eta \ln \gamma_0 t $. This dependence holds
until $t\lesssim \gamma_{\min}^{-1}$, where $\gamma_{\min}$ is the
\emph{minimal} relaxation rate in the system. At $t \gtrsim
\gamma_{\min}^{-1}$ the quantity $\Gamma (t)$ saturates at the value
of the order of $\Gamma_\infty \sim \eta \ln
(\gamma_0/\gamma_{\min})$. The random process in this case is
non-Markovian and cannot be fully characterized by the pair correlation
function~(\ref{eq:12}).

The above estimates do not describe decay of the free induction
signal due to beats between contributions of different
fluctuators. For $t \ll \gamma_0^{-1}$ this decay can be evaluated
in the same way as influence of the static inhomogeneous
broadening~\cite{KlauderAnderson,Laikhtman}.   Averaging
Eq.~(\ref{eq:tmp01}) we get $\cK_f (t)=\Gamma_\infty t$. At large
time, $t \gg \gamma_0^{-1}$, one can assume that the probability
to find a \f in a given state is just the equilibrium one. Thus
$$ \psi_f(t) \approx 1-n_0+n_0e^{i(1-n_0)vt}
=1-n_0\left(1-e^{i(1-n_0)vt}\right)\, .$$
Averaging this solution over the hopping rates and positions of the
fluctuators we arrive at the same result. Consequently, the free
induction signal decays much more rapidly than the phase memory.

The important point is that at large observation time, $t \gg
\gamma_0^{-1}$, there are \emph{optimal} \fs responsible for
decoherence. The distance $r_{\text{opt}}(T)$, between the optimal \fs and
  the qubit is determined by the condition
  \begin{equation}
    \label{eq:aux1}
    v(r_{\text{opt}}) \approx \gamma_0(T)\, .
  \end{equation}
   If the coupling decays as $1/r^b$ and the \fs are distributed in a
   $d$-dimen\-sio\-nal space, then
$r^{d-1}\, dr \to \cP(v) \propto v^{-(1+d/b)}$. {}From this one
concludes that at $d \le b$ the decoherence is controlled by
optimal fluctuators located at the distance $r_{\text{opt}}$
\emph{provided they exist}. At $d >  b$ the decoherence at large
time is determined by most remote fluctuators with $v=v_{\min}$.
If $d \le b$, but the closest \f has $v_{\max} \ll \gamma_0$, then
it is the quantity $v_{\max}$ that determines the decoherence. In
both last cases $\cK(t) \propto t^2$, and one can apply the
results of Ref.~\onlinecite{Paladino}, substituting for $v$ either
$v_{\min}$ or $v_{\max}$.  Since $r_{\text{opt}}$ depends on the
temperature there can exist a specific mesoscopic behavior of the
decoherence rate. A similar mesoscopic behavior of the decoherence
has been discussed for a microwave-irradiated Andreev
interferometer~\cite{Lundin}.

\section{Simulations}

The procedure outlined above leaves several questions unanswered.
First, how many experimental runs one needs to obtain the
ensemble-averaged result for a single fluctuator? Second, when the
contributions of several fluctuators can be described by averages
over the fluctuators' parameters?

 The second question is the most
delicate.  The situation with a qubit interacting with environment
in fact differs from that of a resonant two-level system in spin
or phonon echo experiments. In the first case the experiment is
conducted using a single qubit surrounded by a set of fluctuators
with fixed locations, while in the second case \emph{many}
resonant TLSs participate the absorption. Consequently, one can
assume that each TLS has its own environment and calculate the
properties averaged over positions and transition rates of the
surrounding fluctuators. How many surrounding fluctuators one
needs to replace the set of fluctuators with fixed locations (and
transition rates) by an averaged fluctuating medium?

\subsection{One fluctuator}

\begin{figure}[ht]
{\includegraphics[width=.45\textwidth]{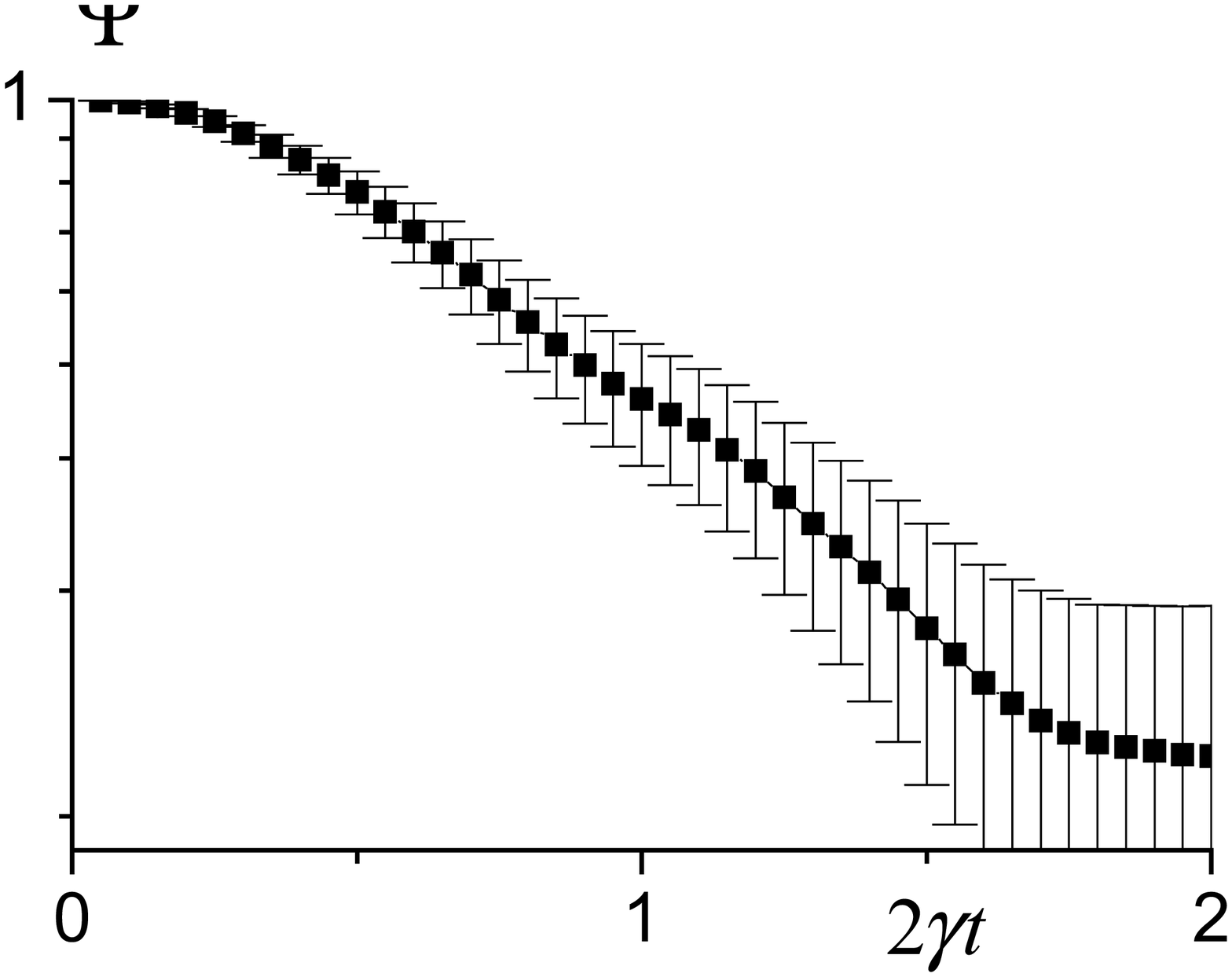} \hfill
\includegraphics[width=.45\textwidth]{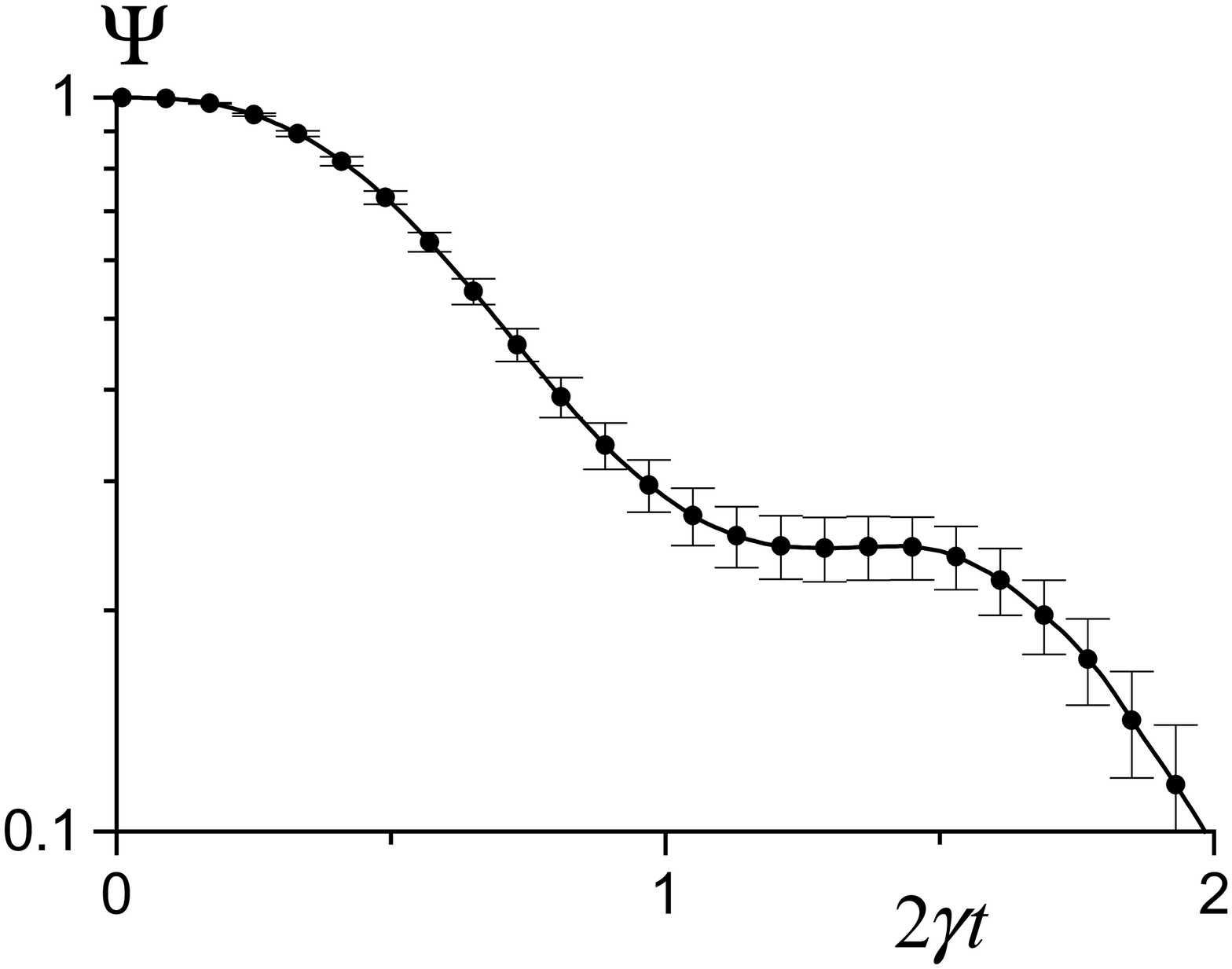}
 \caption{Two-pulse echo decay (in logarithmic
scale) for the case of a single neighboring fluctuator,
$u/2\gamma=5$. The left panel corresponds to 100
realizations of the random switching process, while the right one
corresponds to 1000 realizations. \label{fig:4}}}
\end{figure}

To answer these questions we have made a series of simulations.
First, using Eq.~(\ref{eq:016}) we have calculated the two-pulse echo
decay for a qubit with a single
neighboring fluctuator. The random switching between the $\xi=0$
and $\xi=1$ states of this fluctuator was modeled by the Poisson
process with time constant $\gamma$. The result of simulations for
$u/2\gamma=5$ are shown in Fig.~\ref{fig:4}. The left panel
presents the average over 100 random runs, while the right one
shows the average over 1000 runs. Ultimate averaging over infinite
number of runs would give the analytical result (\ref{eq:04})
shown by the curve 1 in Fig.~\ref{fig:04}.
One can see that averaging over 1000 runs is sufficient to
reproduce the analytical result with good accuracy, in particular,
to observe the characteristic plateau around. Note that the
plateau is qualitatively similar to experimentally observed for
the charge echo in Josephson qubits.~\cite{Nak} Note also that
with only 100 runs made, the dispersion of the signal indicated by
the error bars is huge. Therefore, experimentally one needs at
least many hundred runs to obtain reliable averages.

\subsection{Check of the Holtsmark procedure}

 To check the validity of the summation over different
fluctuators using the Holtsmark procedure, we perform simulations
for many fluctuators. The fluctuators are assumed to be uniformly
distributed in space at distances smaller than some $r_{\max}$.
Then, the normalized distribution function of the coupling
constants and relaxation rates, $\cP(u,\theta)$ can be specified
as
\begin{equation}
\cP(u,\theta)=(\cN u_{\min}/u^2)\, \left[\sin \theta \, \ln (
\tan\theta_{\min}/2 )\right]^{-1}
 \label{psimul2}
\end{equation}
with $u \in \{u_{\min},\infty\}$, while $\theta \in
\{\theta_{\min},\pi/2\}$. Here small $\theta$ correspond to slow
fluctuators, $\gamma=\gamma_0 \sin^2\theta$. The quantity $\eta$
(\ref{eq:06}) that characterizes the fluctuator density is given
now as $\eta=\cN u_{\min}/\ln (\tan\theta_{\min}/2 )$.

\begin{figure}[ht]
\sidebyside {\hspace*{-0.25in}
\includegraphics[width=.5\textwidth]{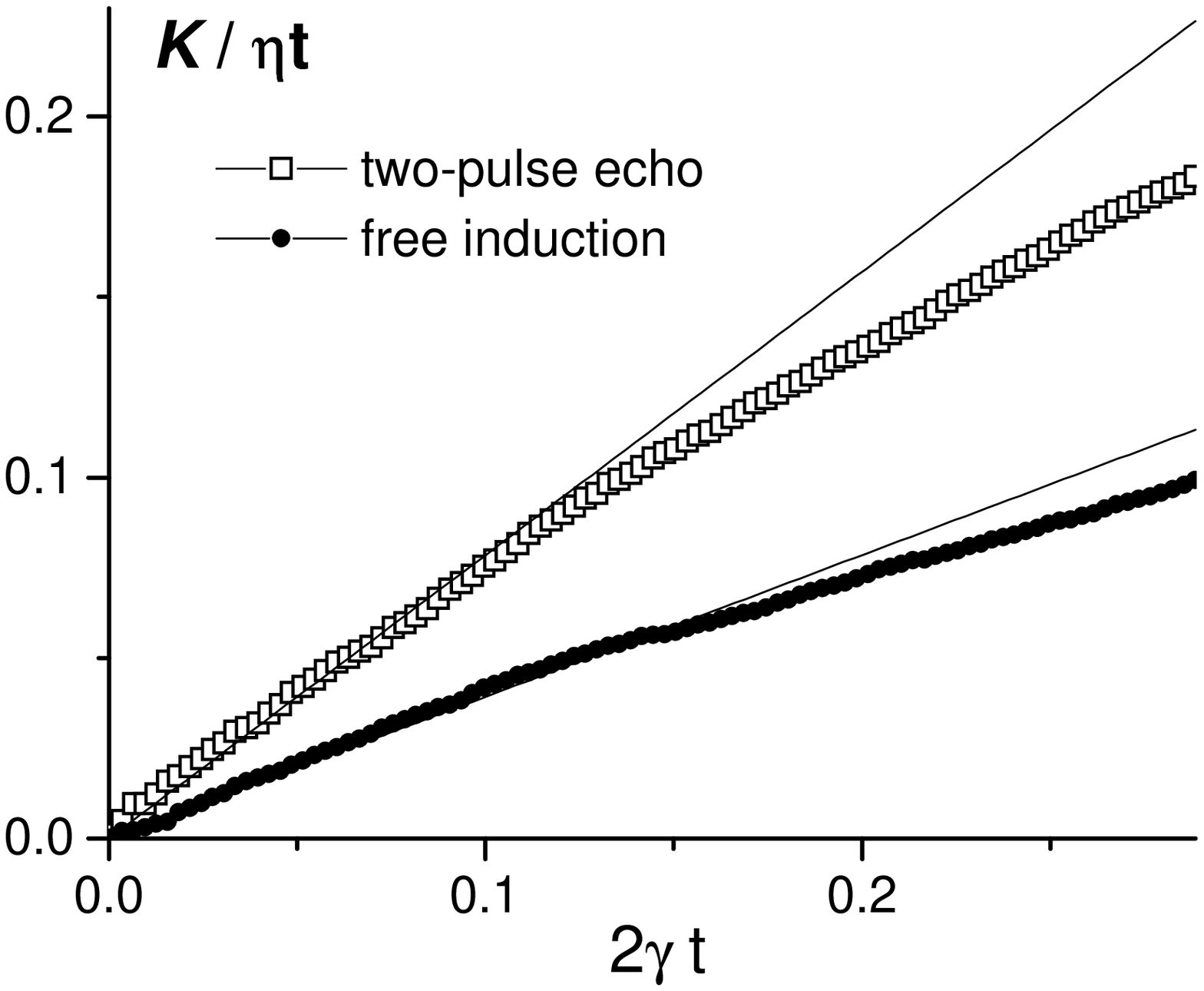}
\caption{Short-time phase memory decay of two-pulse echo (1) and
free induction (2) for $u_{\min}=2\gamma$. Lines (3) and (4),
correspond to analytically calculated slopes: $\pi/4$ and
$\pi/8$, respectively. The results are averaged over 25000
realizations of random telegraph noise in $\cN=10$ fluctuators
randomly distributed in space.
 \label{fig:5}}} {\hspace*{-0.25in}
\includegraphics[width=.5\textwidth]{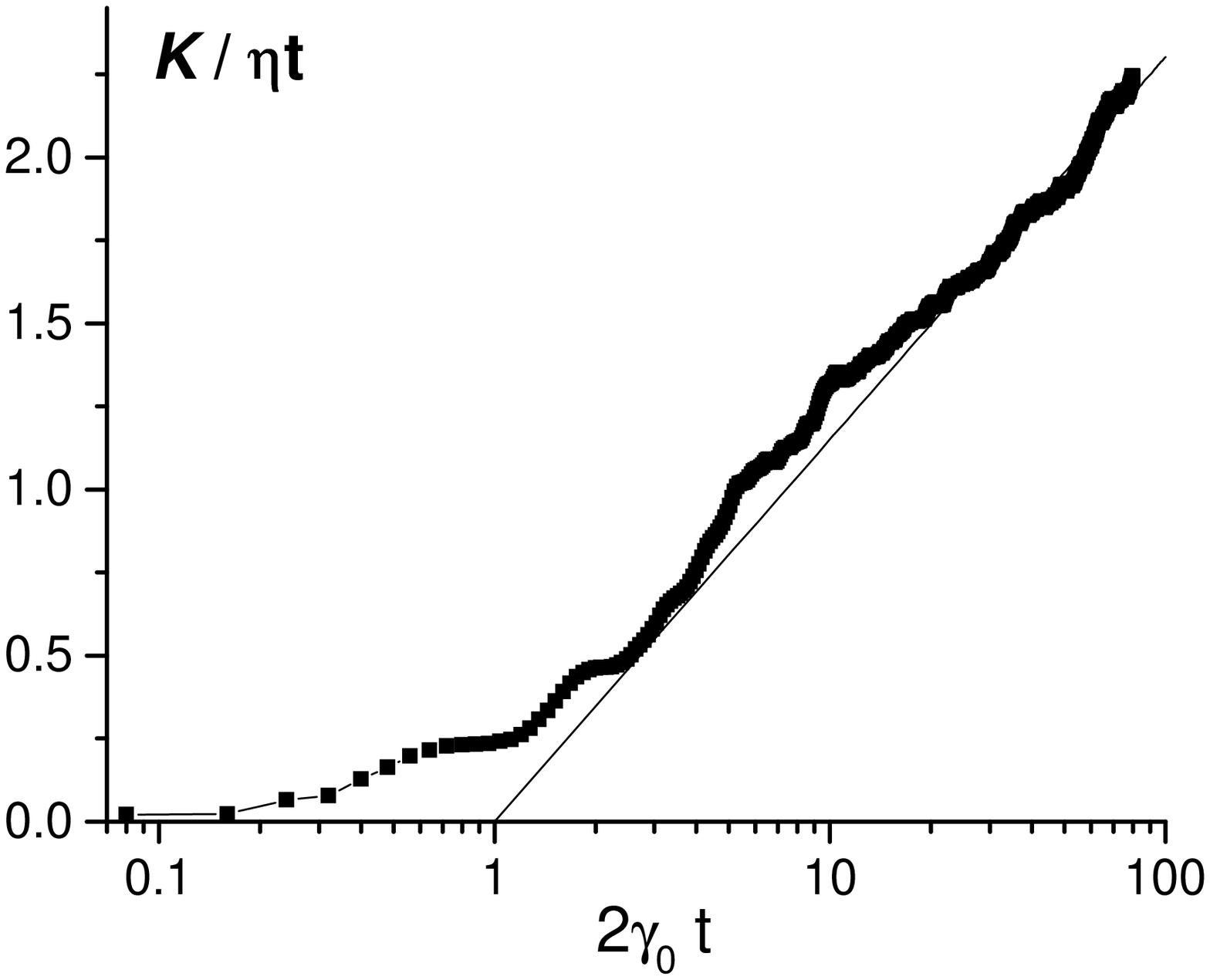} \caption{Long-time
phase memory decay of the free induction for
$u_{\min}=0.02\gamma$. Solid line corresponds to
$(1/2)\ln(2\gamma_0t)$. The results are averaged 
over 5000 realizations of random telegraph noise in $\cN=10$
fluctuators randomly distributed in space. \label{fig:6}}}
\end{figure}

The results of simulations for small times are shown in
Fig.~\ref{fig:5}.  For simplicity, in these simulations the
transition rate was assumed to be the same, $\gamma=\gamma_0$, for
all fluctuators. In order to check the analytical result
(\ref{eq:09}) for the free induction signal, it is convenient to
plot $\cK(t)/\eta t$ versus
$2\gamma t$. One can see that the predicted asymptotic behavior
works rather well for $\gamma t \ll 1$. By substituting
(\ref{eq:04}) into (\ref{eq:07}) it can be easily shown that the
two-pulse echo signal has a similar asymptotic for small $t$, $K
\propto t^2$, only the coefficient is twice that for the free
induction signal. This result is also perfectly reproduced by the
simulations, which justifies the use of the Holtsmark procedure
for $\gamma t \ll 1$, i.~e. when $K$ is small.
 The results for large times are
shown in Fig.~\ref{fig:6}. Here it was important to take into
account scatter in values of $\gamma$ because behavior at large
$t$ is controlled by numerous fluctuators that flip very seldom,
i.~e. have small $\gamma$.  One observes that analytically
predicted behavior of the phase memory for the free induction
signal, $K \propto t \ln t$ at $t \gg \gamma_0^{-1}$ is fully
confirmed by the simulations. The analytical result was obtained
by making a rough cutoff of slow fluctuators at $\theta=(\gamma_0
t)^{-1/2}$ that led to $\cK_{pm}(t) \approx \eta t \ln \gamma_0
t$, see Eq.~\ref{eq:09}. From simulations we can see that a more
accurate expression at large times is $\cK_{pm}(t) \approx (\eta
t/2) \,  \ln 2\gamma_0 t$,  which corresponds to the straight line
in Fig.~\ref{fig:6}.

\subsection{Many fluctuators with fixed locations}

The curves presented in Figs.\ref{fig:5} and \ref{fig:6} were
calculated by averaging over many random sets of fluctuators. This
allowed us to make a reliable check of the analytical results
based on the Holtsmark procedure. The next step is to check
whether it is appropriate to average over the fluctuators
positions though in a real system the fluctuators' parameters are
fixed. For this purpose we compare the results for three different
sets of fixed fluctuators with different coupling constants
distributed again according to Eq.~{(\ref{psimul2}), however with
fixed $\theta$. For each set of fluctuators, we find the average
signal over 1000 runs and show it in Fig.~\ref{fig:7}.
\begin{figure}[ht]
{\includegraphics[width=.5\textwidth]{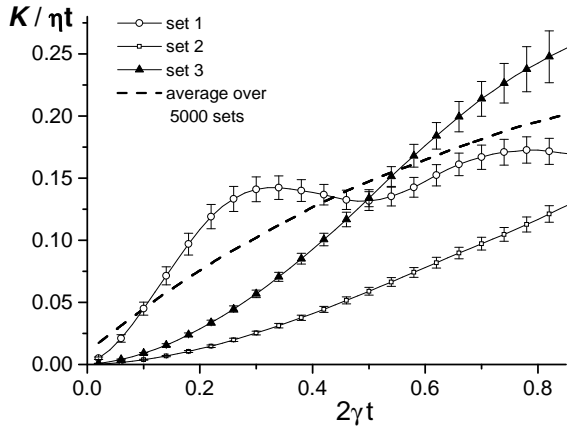} \hfill
\includegraphics[width=.5\textwidth]{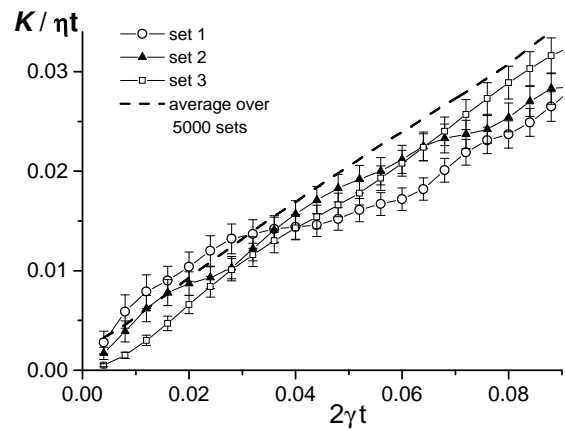} }
\caption{Two-pulse echo decay for three different fluctuators sets
$u_{\min}=2\gamma$, $\cN=10$ (left panel), and 300 (right panel).
 \label{fig:7}}
\end{figure}
 The results strongly depend on the fluctuator density
$\eta=\cN u_{\min} $. For $\eta=300$, the right panel, different
sets of fluctuators lead to similar behavior of $K(t)$, rather
close to the ``expected'' behavior obtained by averaging over 5000
different sets. For larger densities the reproducibility is even
better. However for $\eta=10$, the left panel, each set of
fluctuators is characterized by a very specific type of the
signal. The function $K(t)$ for a given set usually differs
dramatically from the ``expected'' behavior that we obtained by
averaging over 5000 sets. For such a low fluctuator density, it is
hopeless to fit the experimental data with our analytical formulas
obtained by averaging over fluctuator ensemble, like
Eq.(\ref{eq:09}). Note that small error bars on the plot mean only
that the measured signal is reproducible if it is averaged over
1000 experimental runs. However, even a slightly different
arrangement of fluctuators in the gate can produce a very
different signal
\begin{figure}[ht]
\centerline{\includegraphics[width=7cm]{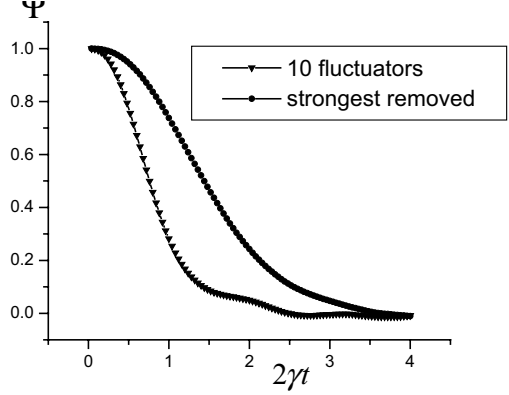} }
 \caption{Two-pulse echo signal for the case of a $\cN=10$ neighboring
 fluctuators with the same transition rate , $\gamma$. It is
 assumed that $u_{\min}=\gamma$.
\label{fig:8}}
\end{figure}

If the fluctuator density is low, or, in other words if we are in
the {\em mesoscopic} limit, the signal should be essentially
determined by one, the most important fluctuator. To check this we
have calculated the echo signal in the presence of $\cN$
fluctuators, and compared it with the signal in the absence of the
strongest fluctuator. The results are shown in Fig.~\ref{fig:8}.
One can see that one needs really many fluctuators to avoid strong
mesoscopic fluctuations. More detailed studies of mesoscopic
fluctuations are planned for future.

\section{Comparison with the noise in the random frequency deviation}
\label{Flicker}

The conventional way is to express the environment-induced
decoherence through the noise spectrum, $\cS_\cX (\omega
)=2\int_{0}^\infty  dt\, e^{i
  \omega t} \left \langle \langle  \cX (t)
  \cX (0)  \rangle_\xi \right \rangle_{\! F}$.
Using Eq.~(\ref{eq:005b}) we get
$$\cS_\cX (\omega )=2\cos^2 \theta_q\left \langle \sum_i
  \frac{2\gamma_i }{\omega^2+(2\gamma_i)^2}\cdot  v_i^2\right
  \rangle_{\! \!F}\, .$$

Let us start averaging over \fs by integration over $\theta_i$. Since
we are interested in small frequencies, we can replace $\sin \theta
\to \theta$, $\cos\theta \to 1$ and replace the upper limit $\pi/2$ of
the integration by infinity. In this way we get
$\cS_\cX (\omega )=\pi\eta  u_{\max}/2\omega$.
Here we have taken into account that the summation over the
fluctuator strength, $u$, is divergent at the upper
  limit corresponding to the \textit{minimal} distance,
  $r_{\min}$, between the fluctuator and the qubit.
 Thus, the closest \fs are most important.
We observe that our model leads to $1/f$ noise in the random force
  acting upon the qubit. However, the noise is mainly determined by
  the \emph{nearest} fluctuators, while decoherence (at long times) is
  dominated by the fluctuators at the distance $r_{\text{opt}}$ given by
  Eq.~(\ref{eq:aux1}). Since $r_{\text{opt}} \gg r_{\min}$,
  the decoherence
  \emph{cannot}, in general, be expressed only through $\cS_\cX
  (\omega )$.

Now we can compare our result given by Eq.~(~\ref{eq:09}). From
Eq.~(\ref{eq:204}) one obtains (cf. with Ref.~\onlinecite{Shnirman})
$$\cK(t)=(1/2) \eta u_{\max}t^2 |\ln \omega_{ir} t|$$
where $\omega_{ir}$ is the so-called intrinsic infrared cutoff
frequency for the $1/f$ noise.\cite{Leggett} It is clear that the
results differ significantly. Even in the case when $\gamma_0 t
\ll 1$ when the random process is Markovian, results differ both
by order of magnitude and temperature dependence. The reason for
this discrepancy is that dephasing  and $1/f$-noise are determined
by different sets of fluctuators.

\section{Applicability range of the model}

Let us discuss the applicability range for the used approach. Firstly,
random fields acting on the qubit were assumed to be
\emph{classical}. This is correct provided the typical hopping rate
of a fluctuator, $\gamma_0$, is much less than its typical interlevel
spacing, $E \sim T$. This is the case, indeed, for \f interaction
with both with phonons and electrons because of weak coupling.
Secondly,
we did not discuss the mechanism of decoherence due to direct translation of
excitation from the qubit to fluctuators. The most important part of
such interaction can be written as
$$\cH_{\text{tr}}=\sum_i U_i \, \left(\sigma_+\tau_-^{(i)}+
  \sigma_+\tau_-^{(i)} \right) \, ,$$
where $\sigma_\pm = \sigma_x \pm i \sigma_y$, and  $\tau_\pm =
  \tau_x \pm i \tau_y$. One can expect that the coupling constant
  $U_i$ is of the same order of magnitude as $v_i$, i. e. $\sim
  g/r^3$. Assuming the constant \f density of states $P_0$ we can
  estimate the typical energy defect for translation of an excitation
  to the fluctuator at the distance $r$ as $ (\delta E) \sim  (Rr^3)^{-1}$.
The effect of the above off-diagonal interaction can be then estimated
  as $U/(\delta E) \approx P_0 g \approx \eta/T$. This ratio should be
  small within the applicability range of our theory. Indeed, qubit
  will not be useful if its characteristic decay rate, $\tau_\varphi
  \approx \eta$, exceeds its
  interlevel spacing $E_0$, which should be, in turn, much less than
  the temperature. However, at very low temperatures and at not too
  long observation times the above processes could be
  important~\cite{Burin}.

  Another issue, which has not been analyzed, it the decoherence near
  the degeneracy  points where $\cos \theta_q \to 0$. These points
  are of specific importance since linear coupling between the
  qubit and the fluctuators vanishes. The conventional way, see,
  e. g., Ref.~\onlinecite{Makhlin2}) is to introduce the model
  coupling as $V_2=\lambda\cX^2(t)\, \sigma_z$. We believe that the
  model still needs a careful derivation.

\section{Conclusions}

The simple model discussed above reproduces essential features of the
decoherence of a qubit by ``slow'' dynamical defects --
fluctuators. This model is valid for qubits of different types.
The main physical picture is very similar to that of the
spectral diffusion in glasses.

The phase memory decay is due to irreversible processes in the \f
system. In the case on ensemble-averaged measurements it can be
directly determined from the echo-type experiments. In the experiments of the
free-induction type the decay of the signal is due both to the finite
phase memory and to the beats between different values of the qubit
eigenfrequency.

The effective rate of the phase memory decay depends on the
relation between the typical interaction strength, $\eta (T)$, and
the typical \f relaxation rate $\gamma_0(T)$. The first quantity
is just a typical deviation of the qubit eigenfrequency produced
by a \emph{thermal} \f (with the inter-level spacing of the order
of temperature $T$)  located at a typical distance.  The second
quantity is the maximal flip-flop rate for the thermal
fluctuators.

The estimates for the  phase memory time in the limiting cases
$\eta \gg \gamma_0$ and  $\eta \ll \gamma_0$ are summarized in
Eq.~(\ref{eq:10}). In the first limiting case during the decoherence
time only few \fs flip. Consequently, the decoherence is governed by
Markovian processes. In the opposite limiting case, typical \fs
experience many flip-flops during the decoherence time. the subsequent
deviations in the qubit frequency being
\emph{statistically-dependent}. As a result, the decoherence process
is essentially non-Markovian.

The details of the decoherence depend strongly on the concrete type of
the fluctuators, namely on the distribution of their flip-flop rates,
on the range of their effective field acting upon the qubit, and on
the distribution of the \fs in real space.

\begin{acknowledgments}
 This research was supported in part by the
National Science Foundation under Grant No. PHY99-07949, by the
International Center for Theoretical Physics, Trieste, Italy, and
by the US DOE Office of Science under contract No.
W-31-109-ENG-38. One of the authors (D.V.S.) is thankful to NorFA
and FUNMAT/UiO for financial support. We are grateful to J.
Bergli, V.~Kozub, V.~Kravtsov, V. Tognetti, and V. Yudson for
discussions of theoretical issues and to Y. Nakamura, J. S. Tsai,
Yu. A. Pashkin, and O. Astafiev for discussions of experimental aspects.
\end{acknowledgments}

\end{document}